\documentclass[12pt]{iopart}
\usepackage{iopams}
\usepackage{epsfig}
\usepackage{epsf}
\def\be{\begin{equation}}
\def\ee{\end{equation}}
\def\bea{\begin{eqnarray}}
\def\eea{\end{eqnarray}}

\begin{document}
\title{Spectral fluctuation characterization of random matrix ensembles through wavelets}

\author{P. Manimaran$^2$, Prasanta K. Panigrahi$^1$\footnote[1]{\texttt{prasanta@prl.res.in}},
 and P. Anantha Lakshmi$^2$}

\address{$^1$ Physical Research Laboratory, Navrangpura, Ahmedabad
380 009, India}

\address{$^2$ School of Physics, University of Hyderabad,
Hyderabad 50046, India}

\begin{abstract}

A recently developed wavelet based approach is employed to
characterize the scaling behavior of spectral fluctuations of
random matrix ensembles, as well as complex atomic systems. Our
study clearly reveals anti-persistent behavior and supports the
Fourier power spectral analysis. It also finds evidence for
multi-fractal nature in the atomic spectra. The multi-resolution
and localization nature of the discrete wavelets ideally
characterizes the fluctuations in these time series, some of which
are not stationary.
\end{abstract}


\date{today}

Recently, a fundamentally different approach to the study of random
matrix ensembles has been taken \cite{rela1,rela2,rela3,rela4}. The
cumulative of the unfolded spectra for a variety of ensembles has
been treated as a time series, on which a Fourier spectral analysis
has been carried out. Very interestingly, this reveals a ubiquitous
$1/f$ power law behavior in the Fourier domain, which has been taken
as the defining characteristic of these random matrix ensembles. In
this context, a number of tools, earlier employed for the analysis
of various time series showing self-similar behavior, can be used
for the characterization of the above data. The well-known methods
like rescaled range analysis \cite{Hurst}, structure function method
\cite{Feder}, wavelet transform modulus maxima \cite{arn1} etc.,
although powerful, have been found to be wanting, when applied to
non-stationary time series. Recently, we have developed a discrete
Daubechies wavelets \cite{daub} based approach \cite{mani1,mani2},
which is well suited to remove local trends in any time series and
faithfully characterize the fluctuations. It is complimentary to the
well studied multi-fractal detrended fluctuation analysis (MFDFA)
\cite{peng,gopi,ple,khu,chen,matia}, which uses appropriate
polynomial fits in local windows for extracting fluctuations. It is
worth mentioning that, the multi-resolution ability of the wavelets
naturally makes them ideal for the analysis of time series showing
self-similar behavior \cite{mandel}.

This letter studies the scaling and correlation behavior and
multi-scaling properties of energy fluctuations in random matrix
ensembles and atomic systems. The primary motivation for this comes
from the aforementioned Fourier spectral analysis, which is known to
yield the Hurst exponent H, with $\alpha= 2H+1$, $\alpha$ being the
exponent of the power law decay.  Since, H is related to the second
moment, it is quite natural to ask the nature of the higher moments,
which determine the multi-fractal characteristic of the time series.
For this purpose, we make use of a discrete wavelet based approach,
developed by us earlier \cite{mani1,mani2}. In this approach, after
removal of the local trend over a given window, the fluctuations are
extracted. The wavelets naturally provide a number of windows of
different sizes, to extract fluctuations at various scales. Using
these, the fluctuation function \cite{xu} is then calculated, which
yields the mono or multi-fractal nature of the time series, when
studied in a log-log plot. We further test the self similar behavior
of the above time series through MFDFA for completeness. Our results
on Hurst exponent matches with Fourier spectral analysis.

It is found that certain atomic systems exhibit multi-fractal
behaviour. It is observed that removing the eigenvalues of the
localized states, which are quantum analogue of classical unstable
orbits, from the energy level data does not alter the nature of the
self-similar behavior, although it affects the Hurst exponent value.
The identification and quantification of localized states are
discussed in Ref.\cite{dilip}. Study of multi-fractal behaviour  has
been carried out earlier in the context of various atoms and ions.
Methods like, box-counting \cite{Saiz} and correlation sum have been
employed \cite{Pawelzik,Cummings}. Keeping in mind the possible
non-stationary nature of the data, as well as efficacy of the
wavelet based approach for extracting fluctuations, we have used the
same here for studying the self-similar nature of the fluctuations.

We give a brief description of the wavelets before proceeding to the
analysis of time series related to various random matrix ensembles
\cite{rand} and atomic level data \cite{santh}. Results and
discussions pertaining to the above analysis are then presented. One
observes mono-fractal behavior for the time series associated with
random matrix ensembles. Two atomic energy level time series show
multi-fractal behavior and another reveals mono-fractal character.
We summarize and conclude in the final section, after pointing out
several directions for future work.

The discrete wavelets provide complete and orthonormal basis
functions, starting from the father wavelet $\phi (t)$ (scaling
function) and mother wavelet $\psi (t)$ [8,21]\cite{daub,primer}.
These functions necessarily satisfy, $\int \phi(t) dt = A$ and $\int
\psi(t) dt = 0$, where $A$ is constant. Scaling and translation of
wavelets lead to $\psi_{j,k} = 2^{j/2} \psi (2^j t - k)$, which obey
the orthogonality conditions: $\langle \phi_{j,k}|\psi_{j,k} \rangle
= 0$ and $\langle \psi_{j,k} | \psi_{j',k} \rangle = \delta_{j,j'}
\delta_{k,k'}$. Any signal belonging to $L^2$ can be written in the
form,
\begin{equation}
\label{cjk_djk}
 f(t) = \sum_{k=-\infty}^{\infty} c_k ~~ \phi_{k}(t)
  + \sum_{k=-\infty}^{\infty}\sum_{j=0}^{\infty} d_{j,k}
~~\psi_{j,k}(t)\,\,\, ,
\end{equation}
where $c_k's$ are the low-pass coefficients and $d_{j,k}'s$ are
the high-pass coefficients. The Daubechies family of wavelets are
made to satisfy vanishing moment conditions: $\int dt~ t^m\psi_{j,
k}(t)=0$. This makes them ideal to isolate polynomial trends from
fluctuations. We make use of the discrete wavelets from Daubechies
family for our analysis of energy level fluctuations.

Consider $E_i, ~ i=1,2,3...,n+1$, the discrete energy levels of
random matrix ensembles or atomic systems, represented as a discrete
time series in the form $E_i$ as shown in Figs. 1 and 2. The trend
and fluctuations from the integrated level density are separated by
the higher order polynomial fitting: $N(E)=N_t(E) + N_f(E)$. The
unfolded energy spectra $\bar{E_i}$ is obtained through the
transformation $N_t(E_i)= \bar{E_i}$. The fluctuations of the energy
level spacings are, $e_i ={\bar{E}_{i+1} - \bar{E}_i}, ~~
i=1,2,3,...,n$, where $<e_i> = 1$. The time series of energy level
fluctuations is given by,

\begin{equation}
\delta_m = \sum_{i=1}^m (e_i - <e_i>); ~~ m=1,...,n.
\end{equation}

The power spectrum of the times series $\delta_m$ is calculated for
various random matrix ensembles and atomic energy levels of Nd, Pm
and Sm. It should be noted that, these have 60, 61 and 62 electrons
respectively. The corresponding number of active valence electrons
are 6, 7 and 8; the complexity increases as the number of active
valence electrons increases. The random matrix ensembles studied are
Gaussian orthogonal ensemble (GOE), Gaussian unitary ensemble (GUE),
Gaussian diagonal ensemble (GDE) and Gaussian symplectic ensemble
(GSE). The cumulative of the unfolded time series are shown in Fig.1
and Fig.2, for the atomic systems and random matrix ensembles
respectively. Regarding the details of the atomic system, interested
readers are referred to Ref.\cite{santh}

Fourier spectral analysis of all the data sets yields, $S(k) \approx
\frac{1}{k^{\alpha}}$, with $\alpha \sim 1$. This is in agreement
with the results of Rela\~{n}o et. al.,, which found that the energy
spectra of quantum systems exhibiting classical chaos are
characterized by $1/f$ behavior. We carried out the spectral
analysis using fast Fourier transform \cite{kor}, and ensemble
averaging was employed in finding the exact slope $\alpha$.

We now proceed to the study of the scaling properties of these time
series through wavelet based fluctuation analysis. We make use of
wavelets from Daubechies family for characterization, since these
naturally remove polynomial trends from data sets. The fractal
nature of the time series is revealed through the study of the
fluctuation function.
In discrete wavelet transform it is well-known that, a given signal
belonging to $L^2$ space can be represented in a nested vector space
spanned by the scaling functions alone. This basic requirement of
multi-resolution analysis can be formally written as \cite{ram},
\begin{equation}
...\subset \nu_{-2} \subset \nu_{-1}\subset \nu_{-0}\subset
\nu_{1}\subset \nu_{2}...\subset L^2,
\end{equation}
with $\nu_{-\infty} = {0}$ and $\nu_{\infty}= L^2$. This provides a
successive approximation of a given signal in terms of low-pass or
approximation coefficients. It is clear that, the space that
contains high resolution signals will also contain signals of lower
resolution. The signal or time series can be approximated at a level
of ones choice, for use in finding the local trend over a desired
window. The fluctuations can then be obtained by subtracting the
above trend from the signal. We have followed this approach for
extracting the fluctuations, by elimination of local polynomial
trends through the Daubechies wavelets.
We compute the fluctuation function in order to ascertain the
self-similar nature of the time series. The $q^{th}$ order
fluctuation function, $F_q(s)$ is obtained by squaring and averaging
fluctuations over all segments:
\begin{equation}
F_q(s) \equiv  \{ \frac {1}{2 M_s} \sum_{b=1}^{2 M_s} [
F^2(b,s)]^{q/2}  \}^{1/q}.
\end{equation}

Here '$q$' is the order of moments that takes any real values. The
above procedure is repeated for variable window sizes for different
value of $q$ (except $q=0$). The scaling behavior is obtained by
analyzing the fluctuation function,
\begin{equation}
F_q(s) \sim s^{h(q)},
\end{equation}
in a logarithmic scale for each value of $q$. If the order $q = 0$
logarithmic averaging has to be employed to find the fluctuation
function:

\begin{equation}
F_0(s) \equiv exp \{ \frac {1}{4 M_s} \sum_{b=1}^{2 M_s} ln [
F^2(b,s)] \}.
\end{equation}

As is well-known, if the time series is mono-fractal, the $h(q)$
values are independent of $q$. For multifractal time series, $h(q)$
values depend on $q$. The correlation behavior is characterized from
the Hurst exponent ($H=h(q=2)$), which varies from $0 < H < 1$. For
long range correlation, $H > 0.5$, $H=0.5$ for uncorrelated and $H
<0.5$ for long range anti-correlated time series.
We refer the interested readers to \cite{mani1,mani2} for the
details of this approach. The power law manifests itself as a
straight line in the log-log plot of $F_q(s)$ versus $s$ for each
value of q:
\begin{equation}
F_q(s) \sim s^{h(q)}.
\end{equation}
For mono-fractal time series, $h(q)$ is constant for all $q$,
whereas for multi-fractal time series h(q) shows non-linear
dependence for all $q$. Here $q$ varies from $-10$ to $+10$. We have
used Db-8 wavelet for capturing fluctuations. The well known Hurst
exponent $H$ equals to $h(q=2)$, which is related to the power
spectral analysis by the relation $\alpha = 2H+1$. Since, the values
of the fluctuations are very small, in order to study the same, we
integrated the time series by subtracting mean. Through this double
integrated time series, the obtained Hurst exponent is $H_1=H+1$.

The Hurst exponent calculated from Fourier analysis compares well
with the wavelet based fluctuation analysis for correlation
behavior. We have also corroborated our results through MFDFA.
\begin{figure}
\centering
\includegraphics[width=3in]{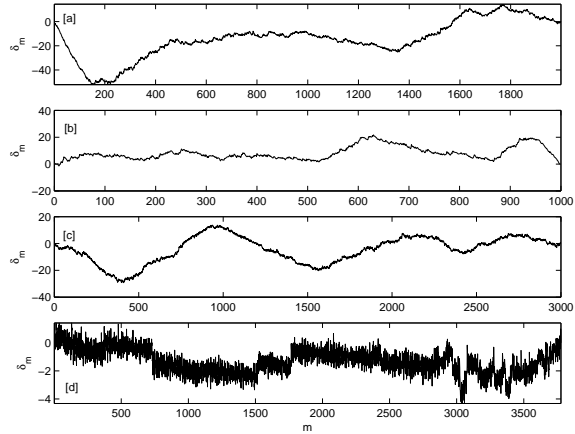}
\caption{Time series of extracted fluctuations from the unfolded
energy spectra of [a] Pm, [b] Nd and [c]and [d] Sm atomic energy
levels.}
\end{figure}

\begin{figure}
\centering
\includegraphics[width=3in]{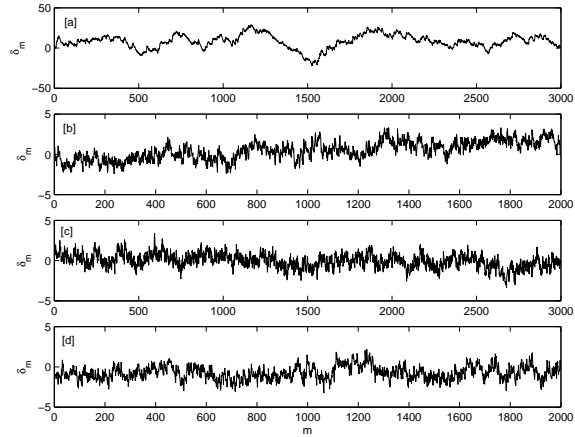}
\caption{Time series of extracted fluctuations from the unfolded
energy spectra of random matrix ensembles [a] GDE, [b] GSE, [c] GOE
and [d] GUE.}
\end{figure}

\begin{table}
\begin{center}
\begin{tabular}{|c|c|c|c|c|c|c|c|c|c|}
  \hline \hline
  q & GDE & GSE & GOE & GUE & Pm & Nd & Sm & $Sm_d$\\
  \hline \hline
-10 &1.5493 &1.1442 &0.9930 &1.0728 &1.4092 &1.5257 &1.1833 & 0.9440\\
-9  &1.5445 &1.1369 &0.9901 &1.0678 &1.4047 &1.5231 &1.1854 & 0.9394\\
-8  &1.5395 &1.1292 &0.9874 &1.0625 &1.4004 &1.5203 &1.1880 & 0.9347\\
-7  &1.5341 &1.1213 &0.9848 &1.0569 &1.3963 &1.5172 &1.1914 & 0.9301\\
-6  &1.5286 &1.1133 &0.9826 &1.0512 &1.3928 &1.5139 &1.1955 & 0.9259\\
-5  &1.5234 &1.1052 &0.9809 &1.0457 &1.3904 &1.5105 &1.2003 & 0.9224\\
-4  &1.5189 &1.0971 &0.9797 &1.0403 &1.3893 &1.5071 &1.2055 & 0.9202\\
-3  &1.5155 &1.0890 &0.9793 &1.0353 &1.3903 &1.5042 &1.2107 & 0.9197\\
-2  &1.5130 &1.0808 &0.9798 &1.0307 &1.3938 &1.5021 &1.2154 & 0.9213\\
-1  &1.5103 &1.0726 &0.9811 &1.0264 &1.3999 &1.5010 &1.2192 & 0.9252\\
0   &1.5056 &1.0644 &0.9830 &1.0222 &1.4080 &1.5006 &1.2215 & 0.9307\\
1   &1.4980 &1.0563 &0.9852 &1.0183 &1.4159 &1.4987 &1.2219 & 0.9365\\
\hline
2   &1.4878 &1.0481 &0.9872 &1.0145 &1.4165 &1.4916 &1.2202 & 0.9409\\
\hline
3   &1.4759 &1.0401 &0.9888 &1.0108 &1.3983 &1.4759 &1.2163 & 0.9436\\
4   &1.4632 &1.0321 &0.9894 &1.0073 &1.3621 &1.4533 &1.2104 & 0.9449\\
5   &1.4507 &1.0244 &0.9892 &1.0039 &1.3251 &1.4289 &1.2028 & 0.9450\\
6   &1.4389 &1.0170 &0.9880 &1.0008 &1.2958 &1.4070 &1.1941 & 0.9444\\
7   &1.4280 &1.0099 &0.9861 &0.9979 &1.2738 &1.3892 &1.1851 & 0.9432\\
8   &1.4183 &1.0032 &0.9836 &0.9951 &1.2571 &1.3751 &1.1763 & 0.9417\\
9   &1.4097 &0.9970 &0.9806 &0.9925 &1.2440 &1.3640 &1.1680 & 0.9401\\
10  &1.4021 &0.9913 &0.9775 &0.9900 &1.2334 &1.3551 &1.1604 & 0.9383\\
  \hline \hline
\end{tabular}
\end{center}
\caption{The $h(q)$ values for different values of $q$ obtained from
wavelet based fluctuation analysis. Here $h(q=2)= H$, is the Hurst
scaling exponent.}

\end{table}

The self-similar behavior of a variety of random matrix ensembles
and atomic level data have been explored through both discrete
wavelets and MFDFA. The scaling behavior corroborates the findings
of Fourier analysis. We found mono-fractal behavior for random
matrix ensemble time series for which computed Hurst exponents
agreed with the results of Relano {\em{et. al}}. Very interestingly,
for one atomic level data, we observed multifractal behavior for Pm
and Nd systems. Sm showed scaling behaviour. Removal of the
eigenvalues of the localized states made the Sm system monofractal
with strong persistence. These results are shown in Table. 1. In the
last column Sm$_d$ represents the atomic level data for Sm when the
eigenvalues of the localized states have been removed. It is worth
noting that, among the three atoms, Sm has the strongest
configuration mixing as it has the largest number of active valence
electrons. This indicates that the lack of sufficient mixing and
effect of localized levels may influence the multifractal nature of
the atomic systems.

We intend to study these aspects carefully in future. The nature of
correlations in other ensembles which appear in various physical
problems also needs investigations. These include embedded Random
Matrix ensembles relevant for finite interacting particle systems
\cite{kota}.

We are thankful to Profs. J. C. Parikh and V.K.B. Kota for many
useful discussions. We acknowledge extensive discussions with Drs.
M.S. Santhanam and D. Angom, to whom we are also thankful for
providing us with atomic level data.

\vskip.5cm {\bf{References}} 
\end{document}